\documentclass{article}

\usepackage[preprint]{neurips_2026}

\usepackage[utf8]{inputenc} 
\usepackage[T1]{fontenc}    
\usepackage{graphicx}       
\graphicspath{{figures/}}
\usepackage{hyperref}       
\usepackage{siunitx}        
\usepackage{url}            
\usepackage{booktabs}       
\usepackage{amsfonts}       
\usepackage{nicefrac}       
\usepackage{microtype}      
\usepackage{xcolor}         
\usepackage{amsmath}        
\usepackage{multirow}       
\usepackage{array}          

\title{Viable Pool Sizing for On-Chain FX Liquidity:\\
Amplification, Capital, and Resilience\thanks{This work was completed at Royal Bank of Canada as part of the RBC Amplify program.}}

\author{%
  Ryan Fang\\
  Royal Bank of Canada\\
  Toronto, ON \\
  \texttt{rfang@fryand.com} \\
  \And
  Ivan Bardziyan \\
  Royal Bank of Canada\\
  Toronto, ON\\
  \texttt{ivanbardziyan@gmail.com}\\
  \AND
  Jessica Wang \\
  Royal Bank of Canada\\
  Toronto, ON\\
  \texttt{jessica.wang255@gmail.com} \\
  \And
  Mayank Anand \\
  Royal Bank of Canada \\
  Toronto, ON \\
  \texttt{mayank.anand@borealisai.com} \\
}

\begin{document}

\maketitle

\begin{abstract}
Financial institutions deploying on-chain FX liquidity face a joint design problem: how much capital to commit, and how to configure the pool, to remain both competitive on trading costs and profitable as a liquidity provider? The StableSwap mechanism~\citep{egorov2020stableswap} interpolates between constant-product (CPMM) and constant-sum (CSMM) market makers~\citep{port2022mixing} via an amplification factor $A$, but neither extreme suits institutional FX: CPMM pools require excessive capital and generate high impermanent loss; CSMM pools are capital-efficient near the peg but drain rapidly under adversarial flow. Using a Merton jump-diffusion price process~\citep{merton1976option} and the loss-versus-rebalancing (LVR) framework~\citep{milionis2022automated}, we map the joint $(A, \mathrm{TVL})$ space to identify configurations that satisfy all three institutional requirements: competitive slippage, positive return, and shock resilience. Minimum viable pool size scales as $\mathrm{TVL}/Q \approx 1000/A$; ROC at that minimum is thin ($\approx 0.054\%$ per horizon) and $A$-invariant; low-$A$ pools ($A \leq 10$) suffer slippage exceeding $200$~bps under a $10\times$ shock, while high-$A$ pools ($A \geq 500$) suffer reserve drain up to $60\%$, establishing both a capital floor and a practical amplification ceiling.
\end{abstract}

\section{Introduction}

Wholesale FX markets clear at sub-pip spreads. An on-chain venue targeting institutional flow must therefore offer slippage within 2~bps, capital at balance-sheet scale, and resilience to the macro shocks (rate decisions, data releases, geopolitical events) that drive large one-sided flows.

No single point on the CPMM--CSMM spectrum satisfies all three. CPMMs, such as Uniswap~v2~\citep{adams2020uniswap}, provide liquidity at any price but at high slippage and impermanent loss. CSMMs offer near-zero slippage near the peg but are fully drainable when the rate moves~\citep{port2022mixing}. StableSwap~\citep{egorov2020stableswap} interpolates between the two via amplification factor $A$, concentrating liquidity near a target rate while retaining a price-recovery mechanism as reserves become imbalanced.

With fee fixed at 1~bp, the institution's design variables are $A$ and total value locked (TVL). The question is not which $A$ optimizes a single metric, but which $(A, \mathrm{TVL})$ pairs are jointly viable. Capital efficiency gains from higher $A$ are offset by shock fragility, and the two constraints bind in different regions of the parameter space.

\section{Related Work}

\citet{adams2020uniswap} established the constant-product AMM baseline; \citet{egorov2020stableswap} introduced StableSwap for near-pegged assets; and \citet{port2022mixing} characterized the CPMM--CSMM tradeoff theoretically. \citet{milionis2022automated} formalized LP costs via the LVR framework, extended to fee-bearing pools by \citet{milionis2023fees}. Uniswap~v3~\citep{adams2021uniswapv3} introduced tick-based concentrated liquidity for general asset pairs; high-$A$ StableSwap is conceptually analogous but purpose-built for near-pegged pairs without manual range management. The FX price process is modeled as a Merton jump-diffusion~\citep{merton1976option}, augmenting GBM with compound Poisson jumps to capture discrete macro shocks. We apply these tools to institutional FX, where slippage tolerances, capital scale, and shock magnitudes differ substantially from the retail DeFi context.

\section{Methodology}

\subsection{Pool Mechanics}\label{sec:pool}

\paragraph{Oracle normalization.}
StableSwap's invariant operates on equal-unit reserves. Since FX rates are not fixed, the pool maintains a reference oracle price $p^* > 0$ and works with \emph{normalized} reserves:
\[
  \tilde{x} = x_{\mathrm{base}}, \qquad \tilde{y} = \frac{x_{\mathrm{quote}}}{p^*}.
\]
Both are in base-currency units. When $p^*$ equals the true cross-rate, $\tilde{x} = \tilde{y}$ at a balanced pool and liquidity concentrates at the correct price. Updating $p^*$ re-centers the invariant instantaneously without touching raw reserves.

\begin{figure}[t]
  \centering
  \includegraphics[width=0.62\linewidth]{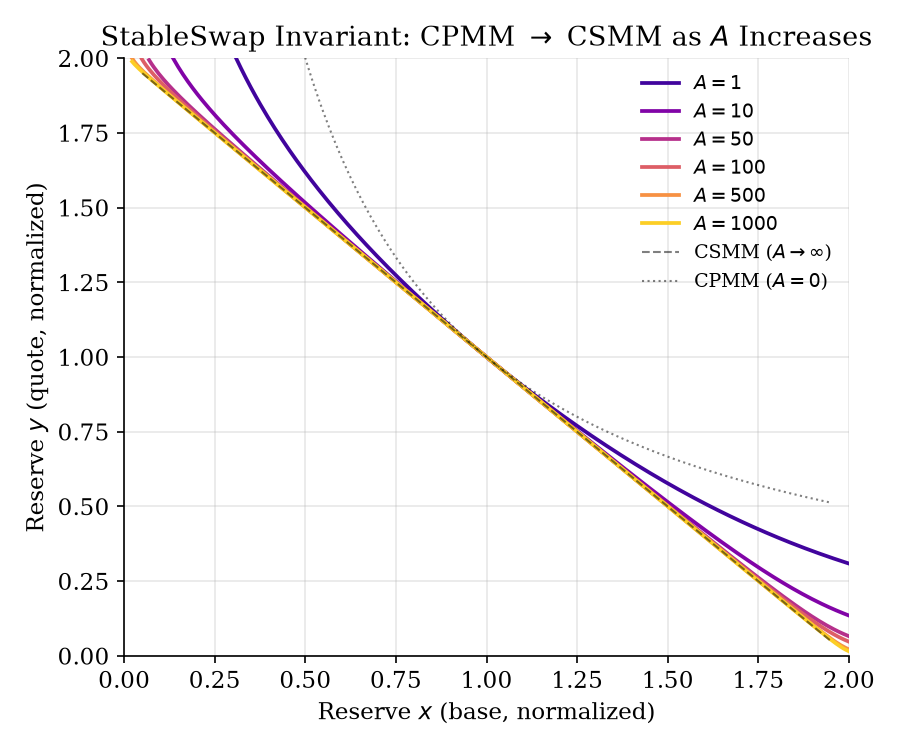}
  \caption{StableSwap invariant curves for $A \in \{1, 10, 50, 100, 500, 1000\}$. Higher $A$ concentrates liquidity near $(D/2, D/2)$, approaching constant-sum behavior (dashed). Lower $A$ spreads liquidity across the full range, approaching constant-product (dotted).}
  \label{fig:invariant}
\end{figure}

\paragraph{Invariant and CPMM--CSMM tradeoff.}
The StableSwap invariant over the normalized reserves $(\tilde{x}, \tilde{y})$ for $n = 2$ is:
\[
  A \cdot n^n \cdot (\tilde{x} + \tilde{y}) + D = A \cdot D \cdot n^n + \frac{D^3}{4\tilde{x}\tilde{y}},
\]
where $D$ is the liquidity invariant. As $A \to 0$: constant-product ($\tilde{x}\tilde{y} = (D/2)^2$); as $A \to \infty$: constant-sum ($\tilde{x} + \tilde{y} = D$). Figure~\ref{fig:invariant} illustrates this geometry. Near balance the curve is nearly linear, producing low price impact; as reserves skew it bends toward constant-product, limiting further drain once a substantial fraction of one reserve is exhausted.

\paragraph{Fees and slippage.}
Each trade of size $Q_t$ pays fee $f \cdot Q_t$ at $f = 10^{-4}$ (1~bp); the net $(1 - f)Q_t$ executes against the invariant. A 1~bp fee matches the tight end of institutional inter-dealer FX spreads and is the protocol-level fee set in StableSwap deployments for stable pairs~\citep{egorov2020stableswap}. Slippage is
\[
  \mathrm{slip}(Q) = \frac{|P^{\mathrm{amm}}_{\mathrm{pre}} - P^{\mathrm{amm}}_{\mathrm{post}}|}{P^{\mathrm{amm}}_{\mathrm{pre}}},
\]
where $P^{\mathrm{amm}}$ is obtained by implicit differentiation of the invariant. With $\mathrm{Ann} = A n^n$:
\[
  P^{\mathrm{amm}} = \frac{\mathrm{Ann} + D^3/(4\tilde{x}^2\tilde{y})}{\mathrm{Ann} + D^3/(4\tilde{x}\tilde{y}^2)} \cdot p^*.
\]
This equals $p^*$ at equilibrium ($\tilde{x} = \tilde{y}$) and diverges as reserves skew. Fee revenue accrues to the institution as sole LP.

\paragraph{TVL/Q scaling.}
Near equilibrium ($\tilde{x} = \tilde{y} = D/2$), implicit differentiation of the invariant gives the second derivative of price with respect to trade size as $\partial^2 P^{\mathrm{amm}} / \partial Q^2 \propto 1/(A \cdot D^2)$. For a fixed slippage target $\epsilon$, the minimum depth scales as $D \propto 1/\sqrt{A\epsilon}$, and since pool depth scales with TVL at fixed $Q$, this yields $\mathrm{TVL}/Q \propto 1/A$. The empirical constant $\approx 1000$ is calibrated from bisection results (Table~\ref{tab:mvt}) and holds within 5\% across all tested $A$ values.

\subsection{Rebalancing, Repegging, and Profitability}

Two mechanisms keep the pool aligned with the off-chain market, and they are triggered independently.

\paragraph{Rebalancing (arbitrage alignment).}\label{sec:rebalancing}
At each step the bot compares $P^{\mathrm{amm}}_t$ to spot $S_t$. If
\[
  |P^{\mathrm{amm}}_t - S_t| > c_{\mathrm{gas}}, \quad c_{\mathrm{gas}} = 5 \times 10^{-7},
\]
the bot executes a price-alignment trade: binary search finds $q^*$ such that $P^{\mathrm{amm}}_t = S_t$ post-trade, capturing the arbitrage spread net of gas cost. This threshold corresponds to a conservative L2 transaction cost of approximately \$0.05 per rebalance on a \$100{,}000 trade, consistent with Layer~2 median fees reported by \citet{ambrosia2026fees}. In practice, total gas burned over 500 steps is $\approx\$0.0003$, compared to fee revenue of $\approx\$5{,}400$ (a ratio of $\sim 10^{-7}$). ROC is therefore insensitive to $c_{\mathrm{gas}}$ across several orders of magnitude; it is determined almost entirely by $f \cdot Q \cdot N_{\mathrm{steps}} / \mathrm{TVL}$.

\paragraph{Repegging (oracle update).}
Repegging updates $p^*$ when the oracle has drifted materially:
\[
  \frac{|p^* - S_t|}{p^*} > \theta_{\mathrm{peg}}, \quad \theta_{\mathrm{peg}} = 0.1\%.
\]
On trigger, $p^* \leftarrow S_t$, re-centering liquidity concentration at the new rate. The 0.1\% threshold is within the typical daily range of G10 FX pairs yet large enough to avoid spurious updates from tick-level noise. Rebalancing corrects the moment-to-moment price gap; repegging corrects structural oracle drift.

\paragraph{Profitability.}
Per~\citet{milionis2022automated}, the arbitrage profit per rebalancing event is $\max(|P^{\mathrm{amm}}_t - S_t| - c_{\mathrm{gas}},\, 0)$. Gross margin over the horizon is:
\[
  \Pi = \underbrace{f \sum_t Q_t}_{\text{fee revenue}} + \underbrace{\sum_t \mathrm{LVR}_t}_{\text{arbitrage capture}} - \underbrace{c_{\mathrm{gas}} \cdot N_{\mathrm{reb}}}_{\text{gas cost}},
\]
where $N_{\mathrm{reb}}$ is the number of rebalancing events. Return on capital is $\mathrm{ROC} = \Pi / \mathrm{TVL}$.

\subsection{Market Simulation}

The off-chain spot $S_t$ follows a Merton jump-diffusion process~\citep{merton1976option}:
\[
  dS_t = \mu S_t\,dt + \sigma S_t\,dW_t + S_t\,dJ_t,
\]
with $S_0 = 1$, $\mu = 0$, $\sigma = 5\%$, $\lambda = 1.5$~yr$^{-1}$, $dt = 1/365$. These parameters are representative of G10 FX pairs under normal conditions: major pairs such as EUR/USD and USD/CAD exhibit annualized realized volatility of 5--8\% and experience roughly one to two significant macro discontinuities per year (central bank surprises, data shocks). The zero-drift assumption ($\mu = 0$) reflects the martingale property of FX rates under risk-neutral pricing. Jumps multiply $S_t$ by $e^Z$, $Z \sim \mathcal{N}(0, 0.02)$ ($\approx \pm 2\%$), calibrated to typical central bank surprise magnitudes. Each step also generates one uninformed trade of size $\approx Q$ in a random direction, representing routine institutional flow.

The resilience experiment uses elevated parameters ($\sigma = 8\%$, $\lambda = 3.0$) to represent a period of sustained macro uncertainty with more frequent shocks. One-sided shocks of $5\times$, $10\times$, and $20\times Q$ are injected at every 50th step. The 50-step interval (50 trading days $\approx$ 10 weeks) approximates the frequency of major scheduled central bank meetings (Federal Reserve, ECB, Bank of Canada), which are the primary source of large one-sided FX flows in G10 markets.

\subsection{Model Assumptions and Viability Metrics}

\paragraph{Core assumptions.}
\begin{itemize}
  \item Institution is the \emph{sole LP}; all fees accrue to it.
  \item Fee fixed at $f = 1$~bp.
  \item Bot has \emph{zero latency}: observes $S_t$ and acts atomically within the same step.
  \item $p^*$ always reflects true spot at repeg time; oracle manipulation not modeled.
  \item Flow is \emph{uninformed}: one i.i.d.\ trade of size $\approx Q$ per step.
  \item $c_{\mathrm{gas}} = 5 \times 10^{-7}$ constant (conservative L2 estimate; see \S\ref{sec:rebalancing}).
\end{itemize}

\paragraph{Viability metrics.}
A pool $(A, \mathrm{TVL})$ is \emph{viable} iff all three hold:
\begin{enumerate}
  \item \textbf{Slippage:} avg.\ $\leq 2$~bps for a $\$100{,}000$ trade.
  \item \textbf{Profitability:} $\Pi > 0$ over 500 steps (\$50M notional traded).
  \item \textbf{Resilience:} under $10\times Q$ shock, max imbalance $< 50\%$ and max slippage $< 100$~bps.
\end{enumerate}
Thresholds follow institutional FX norms: 2~bps is the inter-dealer spread tolerance; 50\% imbalance is the point at which the short reserve becomes operationally thin; 100~bps is a conservative tail-slippage ceiling.

\subsection{Experiment Suite}

(1)~Geometric bisection finds minimum TVL for $\leq 2$~bps slippage at each $A \in \{10, 25, 50, 100, 200, 500, 1000\}$. (2)~A 10-run ensemble sweeps all $(\mathrm{TVL}, A)$ pairs on a $10 \times 7$ log-spaced grid. (3)~A resilience experiment runs 10 ensembles per $(A, \mathrm{TVL})$ under $5\times$, $10\times$, and $20\times$ adversarial shocks. (4)~A sensitivity sweep varies $\sigma \in \{1\%, 3\%, 5\%, 8\%\}$ and $\lambda \in \{0.5, 1.5, 3.0\}$ at $A \in \{50, 100, 200\}$, holding TVL fixed at $\$10$M.

\section{Results}

\subsection{Minimum Viable Pool Size}

\begin{figure}[t]
  \centering
  \begin{minipage}{0.47\linewidth}
    \centering
    \includegraphics[width=\linewidth]{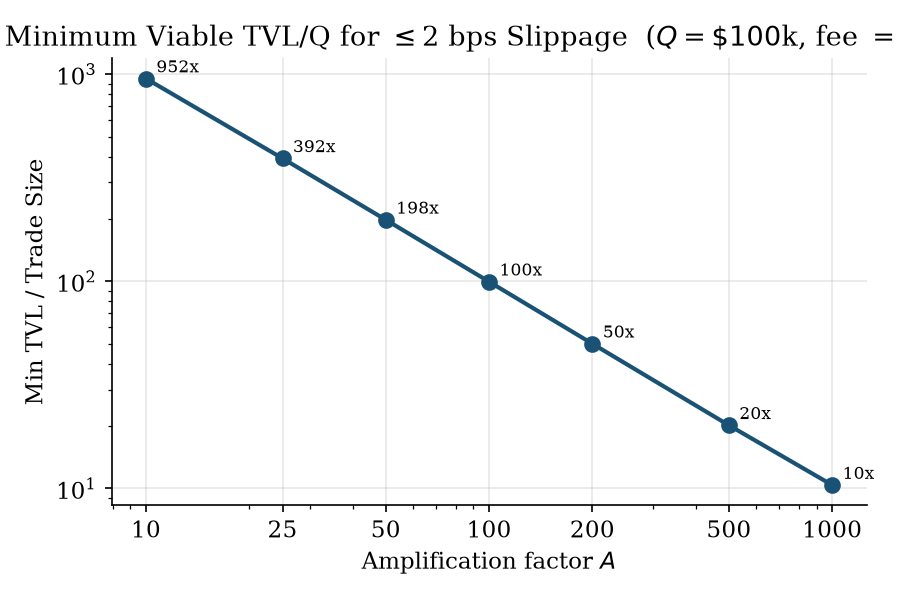}
    \caption{Minimum TVL/Q for 2-bps slippage vs $A$ (log--log). The relationship is approximately TVL/Q $\approx 1000/A$.}
    \label{fig:mvt}
  \end{minipage}\hfill
  \begin{minipage}{0.47\linewidth}
    \centering
    \includegraphics[width=\linewidth]{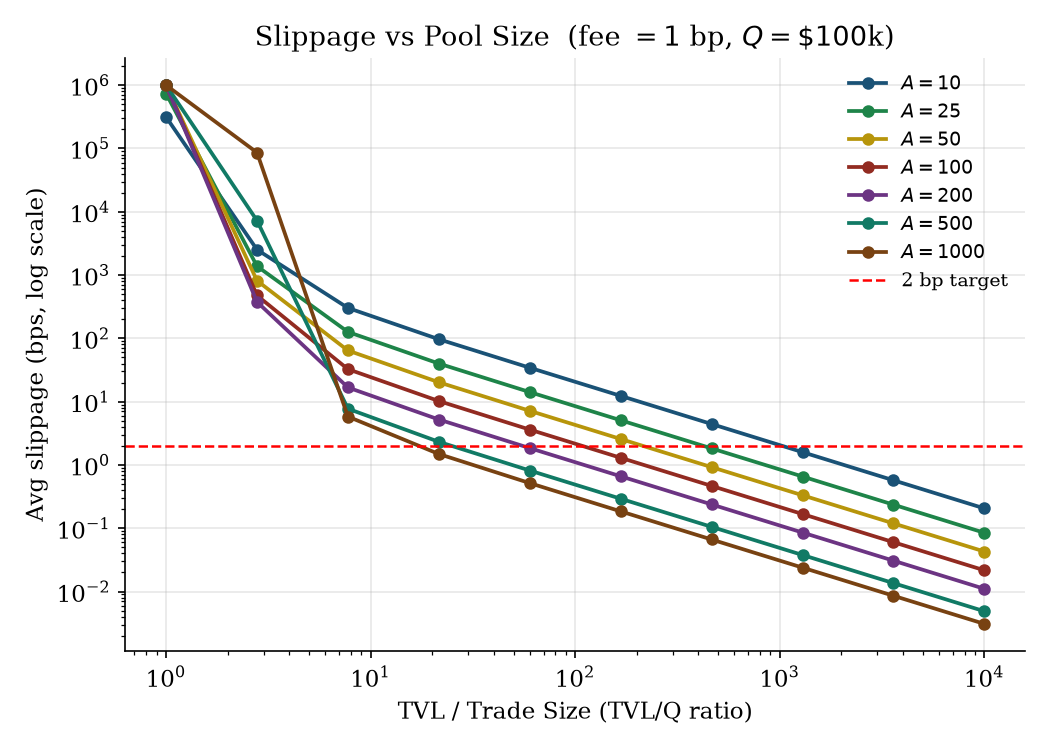}
    \caption{Average slippage (bps) vs TVL/Q ratio across all $A$ values. The 2 bp target (dashed) is met at different TVL/Q thresholds per $A$.}
    \label{fig:slippage}
  \end{minipage}
\end{figure}

Table~\ref{tab:mvt} reports minimum TVL for $\leq 2$~bps slippage at each $A$. The TVL/Q ratio is inversely proportional to $A$: $952\times$ at $A = 10$ ($\approx\$95$M); $10\times$ at $A = 1000$ ($\approx\$1$M). Figure~\ref{fig:mvt} plots this on a log--log scale.

\begin{table}[t]
  \centering
  \caption{Minimum TVL for $\leq 2$ bps slippage per $A$ ($Q = \$100{,}000$, fee $= 1$ bp).}
  \label{tab:mvt}
  \small
  \begin{tabular}{rrrrrrrr}
    \toprule
    $A$ & 10 & 25 & 50 & 100 & 200 & 500 & 1000 \\
    \midrule
    Min TVL & $9.5\times10^7$ & $3.9\times10^7$ & $2.0\times10^7$ & $1.0\times10^7$ & $5.0\times10^6$ & $2.0\times10^6$ & $1.0\times10^6$ \\
    TVL/Q   & 952$\times$ & 392$\times$ & 198$\times$ & 100$\times$ & 50$\times$ & 20$\times$ & 10$\times$ \\
    \bottomrule
  \end{tabular}
\end{table}

The $\mathrm{TVL}/Q \approx 1000/A$ scaling follows from invariant geometry: near equilibrium, marginal price curvature is proportional to $1/A$, so depth required for a fixed slippage target is inversely proportional to $A$. This approximation holds only near equilibrium; the relationship breaks down under large shocks.

\subsection{Return on Capital}

ROC falls monotonically with pool size: fee revenue is set by trade volume ($Q$ fixed), so a larger TVL denominator purely dilutes return. At minimum viable TVL for $A = 100$ ($\approx\$10$M), ROC $\approx 0.054\%$ per horizon.

ROC curves for all $A$ values nearly coincide when plotted against TVL/Q: at the same TVL/Q ratio, pools with $A = 100$ and $A = 1000$ yield identical ROC ($\approx 0.054\%$ at TVL/Q $= 100\times$). Higher $A$ reduces the capital required to reach a given TVL/Q without improving yield; the tradeoff is entirely in resilience.

\begin{figure}[t]
  \centering
  \includegraphics[width=0.62\linewidth]{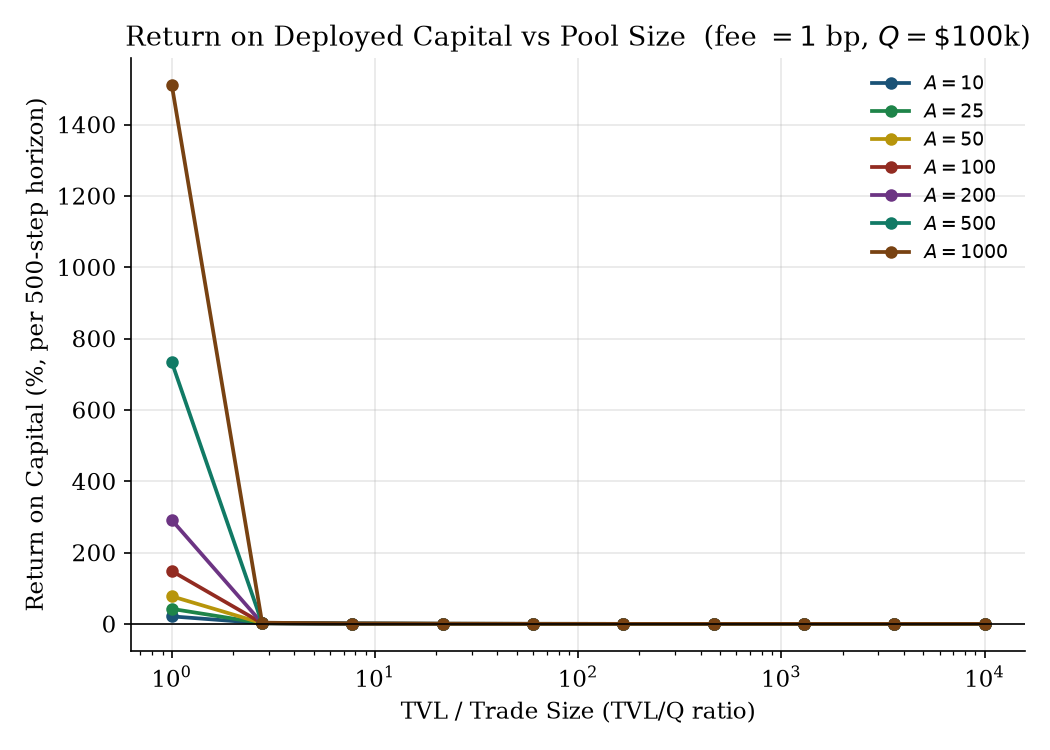}
  \caption{Return on capital (\% per 500-step horizon) vs TVL/Q ratio across $A$ values. ROC is positive everywhere but falls rapidly with pool size. Curves for different $A$ nearly overlap, confirming that profitability is determined by TVL/Q, not $A$ alone.}
  \label{fig:roc}
\end{figure}

\subsection{Resilience Under Adversarial Shocks}

Table~\ref{tab:resilience} reports resilience metrics under a $10\times Q$ shock (\$1M trade, \$10M pool).

\begin{table}[t]
  \centering
  \caption{Resilience metrics under $10\times Q$ adversarial shock (TVL $= \$10$M, 10 runs each). All 10 injected shocks trigger an LVR leakage event regardless of $A$. Max imbalance $= |\tilde{x} - \tilde{y}|/D$; avg imbalance is the mean over all steps; values above $0.5$ indicate over half the pool drained to one side.}
  \label{tab:resilience}
  \small
  \setlength{\tabcolsep}{6pt}
  \begin{tabular}{lrrrrrrr}
    \toprule
    & \multicolumn{7}{c}{Amplification factor $A$} \\
    \cmidrule(lr){2-8}
    Metric & 10 & 25 & 50 & 100 & 200 & 500 & 1000 \\
    \midrule
    Max slippage (bps) & 210 & 88 & 45 & 23 & 13 & 8 & 8 \\
    Max imbalance      & 0.22 & 0.23 & 0.24 & 0.25 & 0.28 & 0.44 & 0.59 \\
    Avg imbalance      & 0.01 & 0.01 & 0.02 & 0.02 & 0.03 & 0.05 & 0.07 \\
    \bottomrule
  \end{tabular}
\end{table}

Max slippage falls with $A$ to $A = 200$ ($13$~bps) and then plateaus: both $A = 500$ and $A = 1000$ yield an $8$~bps mean, though with high-variance tail runs. This plateau reflects the CSMM transition: high-$A$ pools have low price impact near balance but face unbounded impact once pushed past the invariant's efficient range. The unboundedness is a direct consequence of the invariant's asymptotic structure: as one normalized reserve $\tilde{x} \to 0$, the marginal price $P^{\mathrm{amm}} \propto D^3/(4\tilde{x}^2\tilde{y}) \to \infty$ (see the formula in \S\ref{sec:pool}). The curve asymptotes to the axes rather than crossing them, so a sufficiently large one-sided trade always encounters slippage growing without bound. This property prevents complete pool drain at the cost of extreme price impact in the tail.

\begin{figure}[t]
  \centering
  \includegraphics[width=0.72\linewidth]{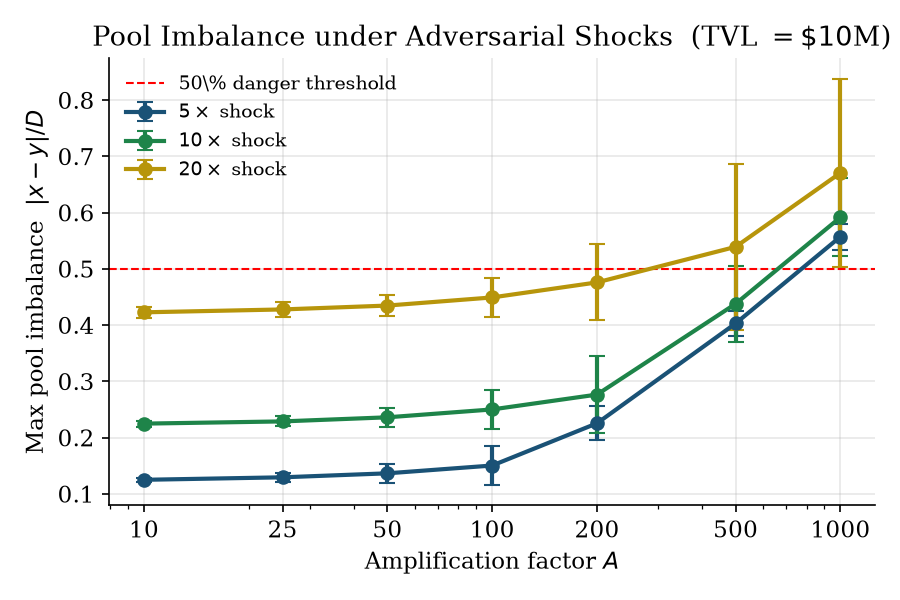}
  \caption{Max pool imbalance under $5\times$, $10\times$, and $20\times$ adversarial shocks vs $A$ (TVL $= \$10$M). Imbalance grows sharply above $A = 200$ for all shock levels, crossing the $50\%$ danger threshold near $A = 500$.}
  \label{fig:imbalance}
\end{figure}

\begin{figure}[t]
  \centering
  \includegraphics[width=\linewidth]{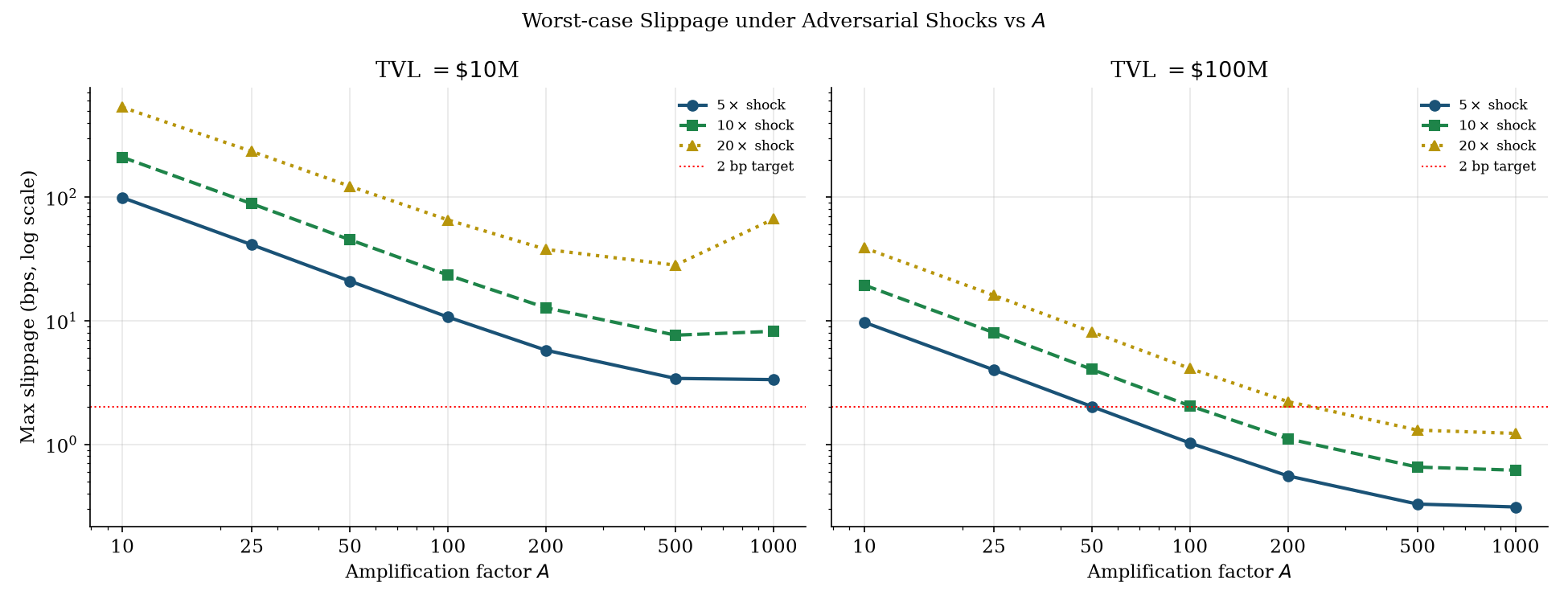}
  \caption{Max slippage under $5\times$, $10\times$, and $20\times$ shocks vs $A$ for TVL $= \$10$M (left) and $\$100$M (right). Slippage falls with $A$ then plateaus as high-$A$ pools transition toward constant-sum behavior, with tail-run variance increasing.}
  \label{fig:shock_slip}
\end{figure}

Imbalance corroborates this: $0.28$ at $A = 200$, $0.44$ at $A = 500$, $0.59$ at $A = 1000$. A pool drained nearly 60\% to one side is vulnerable to complete drain under continued one-sided flow. Figures~\ref{fig:imbalance} and~\ref{fig:shock_slip} show all three shock levels.

\subsection{Sensitivity to Market Conditions}

Rebalancing frequency scales with both $\sigma$ (larger per-step deviations) and $\lambda$ (more frequent jumps), as either increases the rate at which $|P^{\mathrm{amm}}_t - S_t|$ crosses $c_{\mathrm{gas}}$. Curves for $A \in \{50, 100, 200\}$ are nearly indistinguishable, confirming that operational cost is market-driven, not architecture-driven.

\begin{figure}[t]
  \centering
  \includegraphics[width=\linewidth]{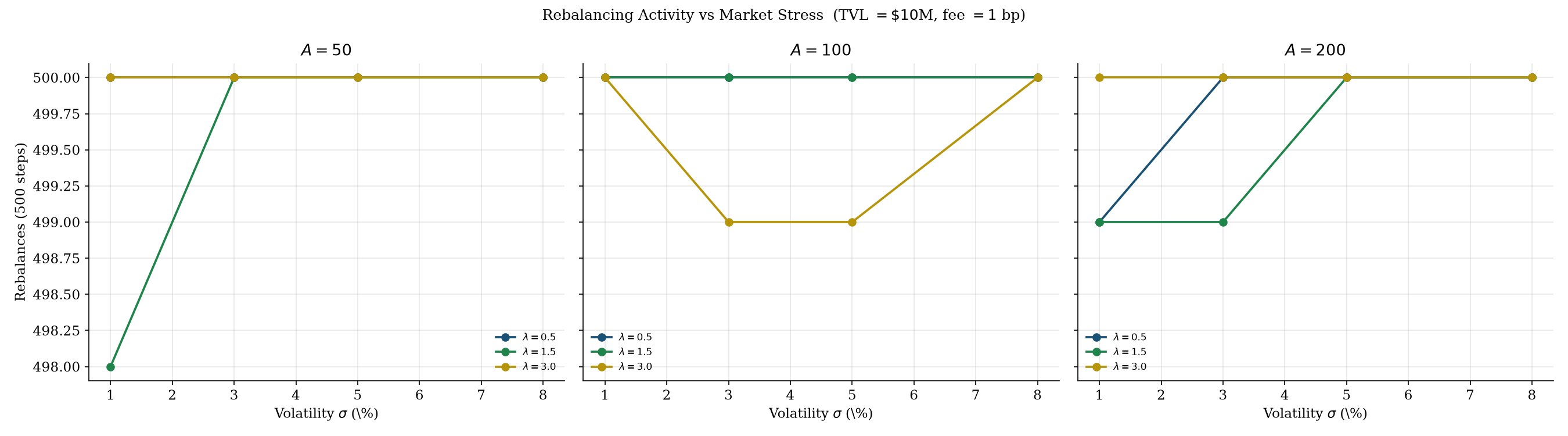}
  \caption{Rebalancing activity vs volatility $\sigma$ for three jump intensities $\lambda$, at $A \in \{50, 100, 200\}$. Curves are nearly identical across $A$ values, indicating that operational burden is market-driven, not architecture-driven.}
  \label{fig:sensitivity}
\end{figure}

\section{Discussion}

\paragraph{Viable region.}
Slippage requires $\mathrm{TVL}/Q \geq 1000/A$. Resilience imposes a ceiling near $A = 200$: above it, max imbalance under a $10\times$ shock reaches $28\%$ at $A = 200$, $44\%$ at $A = 500$, and $59\%$ at $A = 1000$, at which point the short-leg reserve is operationally thin. The intersection places the viable range at $A \in [100, 200]$, requiring $\$5$M--$\$10$M TVL for $Q = \$100{,}000$.

\paragraph{The CSMM ceiling.}
Near balance, the StableSwap curve is nearly linear; near the edges it is nearly vertical and price impact is unbounded. A large unidirectional trade can reach the vertical region before the rebalancer fires, after which further trades extract value at the pool's expense. At $A = 1000$ with TVL $= \$10$M, a single \$1M shock triggers this in the majority of runs.

\paragraph{Profitability requires volume.}
At minimum viable size, 1~bp fee revenue yields $\approx\$5{,}400$ gross margin on a $\$10$M pool per horizon ($0.054\%$). LVR adds a small increment. Revenue scales with trade frequency and notional, not per-trade margin; insufficient order flow cannot justify the locked capital regardless of $A$.

\paragraph{Practical sizing.}
For $Q = \$100{,}000$ at 2-bps slippage: $A \in [100, 200]$, TVL $\in [\$5\mathrm{M}, \$10\mathrm{M}]$. At $A = 500$ capital halves but resilience fails. At $A = 10$ the pool is shock-proof but requires $\$95$M TVL, which is impractical for most institutional balance sheets.

\section{Limitations}

\paragraph{Single LP.}
A permissionless pool may attract passive LPs, diluting fee revenue but also absorbing shock imbalance. Results are conservative on capital but may not hold when LP dynamics are endogenous.

\paragraph{Zero-latency rebalancing.}
In practice, block delays and mempool competition mean the bot may arrive late, allowing external arbitrageurs to extract LVR first and increasing realized slippage.

\paragraph{Constant gas cost.}
We model $c_{\mathrm{gas}}$ as fixed at $5 \times 10^{-7}$~\citep{ambrosia2026fees}. Because gas cost is negligible relative to fee revenue (see \S\ref{sec:rebalancing}), ROC is not materially affected by this choice. However, gas prices are demand-driven and can spike by orders of magnitude during network congestion. At sufficiently high gas cost, rebalancing becomes unprofitable and the bot ceases to fire, allowing the pool to drift from spot and exposing LPs to adverse selection. Dynamic gas modeling is left as future work.

\paragraph{Simplified flow model.}
Real institutional flow exhibits autocorrelation, order fragmentation, and macro-driven informed flow, each of which would affect slippage and profitability.

\paragraph{FX price process.}
The Merton jump-diffusion model is tractable and captures the two key features of FX dynamics: continuous diffusion and discrete macro jumps. However, it assumes constant volatility and independent jumps, whereas FX volatility is well-documented to cluster (GARCH effects) and jumps may be correlated with liquidity conditions. More sophisticated alternatives, such as Heston stochastic volatility or SABR, may yield more accurate calibrations for specific currency pairs. Comparison across price process specifications is left as future work.

\paragraph{Periodic adversarial shocks.}
A sophisticated adversary would time shocks to oracle-lag or low-liquidity windows, potentially causing more severe drain. Reported resilience metrics are therefore lower bounds on vulnerability.

\paragraph{Static oracle.}
Oracle manipulation, latency, and failure are not modeled. A compromised oracle could drain the pool by shifting $p^*$ adversarially.

\section{Conclusion}

On-chain FX liquidity via a concentrated AMM is viable within a well-defined $(A, \mathrm{TVL})$ region. Minimum pool size scales as $\mathrm{TVL}/Q \approx 1000/A$; ROC at that minimum is thin ($\approx 0.054\%$) but positive and $A$-invariant. The binding constraint at high $A$ is not capital efficiency but shock resilience: $A \geq 500$ pools suffer reserve drain exceeding $40\%$ and elevated tail slippage under adversarial flow. For institutional deployments at $Q = \$100{,}000$ and 2-bps slippage, the viable range is $A \in [100, 200]$ with TVL $\in [\$5\mathrm{M}, \$10\mathrm{M}]$.

\bibliographystyle{plainnat}
\bibliography{references}

\clearpage
\section*{Glossary}

\begin{tabular}{>{}p{0.22\textwidth} p{0.72\textwidth}}
  \toprule
  \textbf{Term} & \textbf{Definition} \\
  \midrule
  AMM & Automated Market Maker. A smart-contract-based exchange that uses a mathematical invariant to set prices and execute trades without an order book. \\
  \addlinespace
  CPMM & Constant-Product Market Maker ($xy = k$). High price impact, but the pool never fully drains. \\
  \addlinespace
  CSMM & Constant-Sum Market Maker ($x + y = k$). Near-zero slippage near the peg, but fully drainable under one-sided flow. \\
  \addlinespace
  StableSwap & The AMM invariant (Curve Finance, 2020) that interpolates between CPMM and CSMM via amplification factor $A$. \\
  \addlinespace
  $A$ & Amplification factor. Higher $A$ concentrates liquidity near the peg (CSMM-like); lower $A$ spreads it across the full range (CPMM-like). \\
  \addlinespace
  TVL & Total Value Locked. The aggregate USD value of assets in the pool, and the primary capital sizing variable. \\
  \addlinespace
  $D$ & The StableSwap liquidity invariant (scalar, normalized reserve units), solved via Newton's method at each step. \\
  \addlinespace
  Oracle price ($p^*$) & Off-chain reference FX rate used to normalize reserves. Updated (repegged) when drift from spot exceeds $\theta_{\mathrm{peg}} = 0.1\%$. \\
  \addlinespace
  Rebalancing & Bot action triggered when $|P^{\mathrm{amm}}_t - S_t| > c_{\mathrm{gas}}$. Executes a price-alignment trade $q^*$ to restore $P^{\mathrm{amm}}_t = S_t$, capturing the arbitrage spread. \\
  \addlinespace
  Repegging & Oracle update ($p^* \leftarrow S_t$) triggered independently when oracle drift exceeds $\theta_{\mathrm{peg}}$. Re-centers the pool's liquidity concentration without executing a trade. \\
  \addlinespace
  LVR & Loss-Versus-Rebalancing. Per-rebalancing arbitrage profit equal to $\max(|P^{\mathrm{amm}}_t - S_t| - c_{\mathrm{gas}},\, 0)$. \\
  \addlinespace
  ROC & Return on Capital. Net gross margin $\Pi$ divided by TVL over the simulation horizon. \\
  \addlinespace
  Pool imbalance & $|\tilde{x} - \tilde{y}|/D$. Fractional deviation of normalized reserves from 50/50. A value of 0.5 means one reserve is exhausted. \\
  \addlinespace
  Merton jump-diffusion & Geometric Brownian motion augmented with a compound Poisson jump process, used here to model discrete macro shocks in FX rates. \\
  \bottomrule
\end{tabular}

\end{document}